\documentclass[journal,comsoc]{IEEEtran}
\usepackage[T1]{fontenc}% optional T1 font encoding
\usepackage{cite}

\ifCLASSINFOpdf
   \usepackage[pdftex]{graphicx}
\else
   \usepackage[dvips]{graphicx}
\fi

\usepackage{bm}
\usepackage{amsmath}
\usepackage[cmintegrals]{newtxmath}

\usepackage[numbers]{natbib}
\usepackage{booktabs}
\usepackage{multirow}
\usepackage{multicol}
\usepackage{amsmath}

\usepackage{algorithm}
\usepackage{algorithmicx}
\usepackage{algpseudocode} % defines \For ... \EndFor, \If ... \EndIf, \State, \Return, \Comment

\usepackage{mathtools}
\usepackage{hyperref}

\usepackage{tikz}
\usetikzlibrary{positioning}  % for right=of syntax

\mathchardef\mhyphen="2D

\begin{document}
% \title{Scope expansion for LLM to increase the probability of the correctness of output}
% \title{Internal method extraction to help LLM to learn and think by methods}
\title{Difference-Guided Reasoning: A Temporal–Spatial Framework for Large Language Models}

\author{Hong~Su
% <-this % stops a space
\IEEEcompsocitemizethanks{\IEEEcompsocthanksitem H. Su is with the School of Computer Science, Chengdu University of Information Technology, Chengdu, China.\\
 E-mail: suguest@126.com. \\
\protect\\
}% <-this % stops an unwanted space
\thanks{}}

\markboth{Journal of \LaTeX\ Class Files,~Vol.~14, No.~8, August~2015}%
{Shell \MakeLowercase{\textit{et al.}}: Bare Demo of IEEEtran.cls for IEEE Communications Society Journals}
% * Note that you probably will NOT want to include the author's *
% * name in the headers of peer review papers.                   *
% You can use \IFCLASSOPTIONpeerreview for conditional compilation here if

\maketitle

\begin{abstract}
Large Language Models (LLMs) are important tools for reasoning and problem-solving, while they often operate passively, answering questions without actively discovering new ones. This limitation reduces their ability to simulate human-like thinking, where noticing differences is a key trigger for reasoning. Thus, in this paper we propose a \textbf{difference-guided reasoning framework}, which enables LLMs to identify and act upon changes across time and space. The model formalizes differences through feature extraction, prioritizes the most impactful and latest changes, and links them to appropriate actions. We further extend the framework with mechanisms for abnormal behavior detection and the integration of external information from users or sensors, ensuring more reliable and grounded reasoning. Verification results show that prompting LLMs with differences improves focus on critical issues, leading to higher alignment with desired reasoning outcomes compared to direct prompting. 
\end{abstract}


\IEEEpeerreviewmaketitle

\section{Introduction}
Large Language Models (LLMs) have achieved remarkable progress in recent years and are now widely applied in domains such as question answering \cite{allemang2024increasing}, programming assistance \cite{kazemitabaar2024codeaid}, education \cite{neumann2024llm}, and scientific discovery \cite{zimmermann202532}.  
Their success comes from training on massive corpora and leveraging powerful architectures to predict the next most likely tokens.  
However, despite their breadth of knowledge and fluency, current LLMs primarily operate in a \textbf{reactive manner}: they respond to prompts but do not actively seek differences, anomalies, or new questions in the way humans naturally do \cite{su2025llm}.  

In human cognition, noticing differences plays a central role in reasoning and decision-making.  
A person first identifies a difference or abnormal behavior in the environment, and then decides whether to maintain the current state or take corrective action.  
LLMs, by contrast, rely on statistical next-word prediction, which may overlook subtle but critical differences in states, contexts, or environments.  
This limitation constrains their ability to think dynamically, adapt to evolving situations, and exhibit human-like logical reasoning.  

\begin{figure}[h]
    \centering
    \includegraphics[width=3.2in]{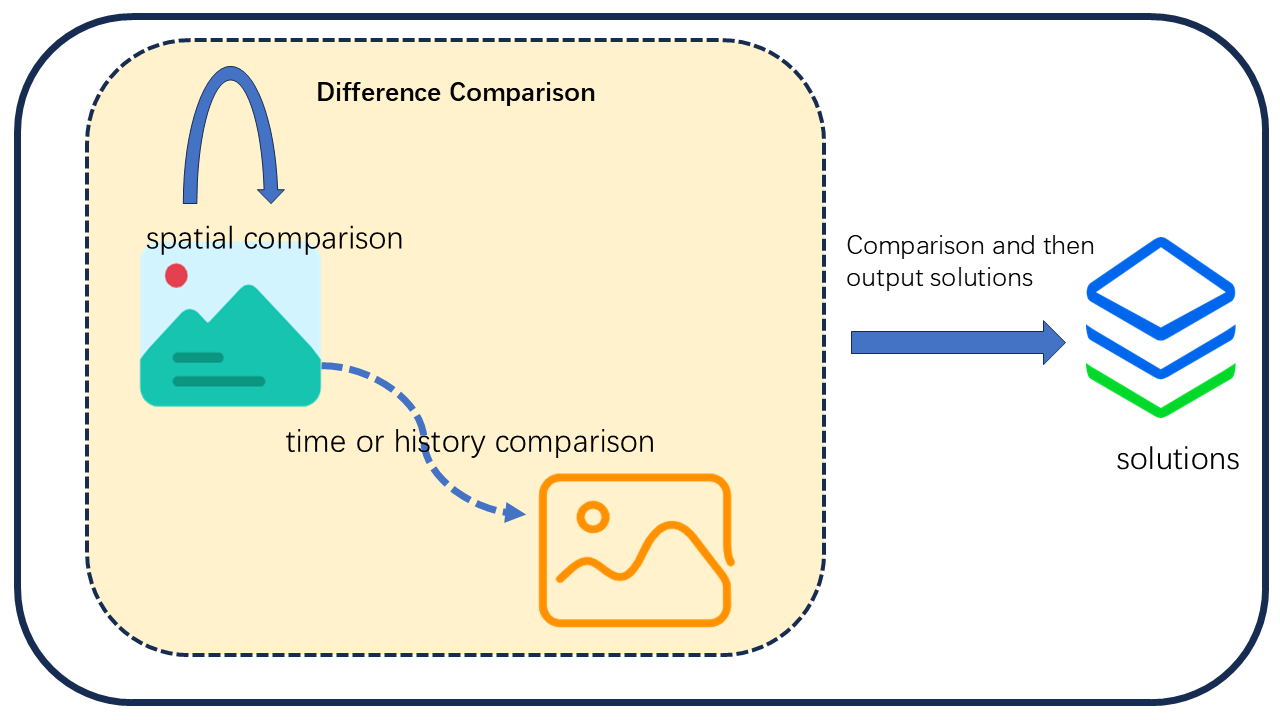}
    \caption{Conceptual illustration of the proposed difference-guided reasoning framework, where detected differences are linked to subsequent reasoning and actions.}
    \label{fig_differenceDemo}
\end{figure}

To address this limitation, we propose a \textbf{difference-guided reasoning framework} for LLMs.  
The core idea is to explicitly guide the model to detect differences—across time, space, or features—and then associate those differences with corresponding actions.  
This process mirrors human reasoning: one first identifies what has changed, and then determines the appropriate response.  
By integrating this mechanism, LLMs can transition from passive response generation to more active and adaptive reasoning.  
An illustration of the framework is provided in Figure~\ref{fig_differenceDemo}.  

The main contributions of this paper are as follows:
\begin{enumerate}
    \item \textbf{Difference-guided reasoning for LLMs.} We propose a framework that enables LLMs to actively detect differences as the entry point for reasoning, thereby moving beyond passive response generation.  
    \item \textbf{Temporal difference reasoning.} We formalize how LLMs can analyze changes across time, identify the latest differences, and link them to appropriate reasoning steps and actions.  
    \item \textbf{Spatial difference reasoning.} We extend the framework to handle differences across subcomponents of complex objects, allowing LLMs to reason about structural or distributional variability within a system.  
\end{enumerate}

The remainder of this paper is organized as follows.  
Section~\ref{sec_related_work} reviews related work.  
Section~\ref{sec_model} presents the proposed difference-guided reasoning model.  
Section~\ref{sec_verification} reports experimental verification and analysis.  
Finally, Section~\ref{sec_conclusion} concludes with a summary and outlines future work.

\section{Related Work} \label{sec_related_work}
Research on enhancing the reasoning ability of Large Language Models (LLMs) has followed several directions.  

\subsection{Prompt Engineering.}
One line of research focuses on \textbf{prompt engineering}, where carefully designed instructions or chains of prompts are used to guide LLMs toward better reasoning.  
For instance, chain-of-thought prompting encourages models to generate intermediate reasoning steps, thereby improving their accuracy on complex, multi-step problems \cite{wei2022chain}.  
This approach has shown strong performance in arithmetic reasoning, symbolic manipulation, and commonsense tasks, where decomposing a problem into smaller steps mirrors human cognitive strategies.  
Follow-up methods, such as least-to-most prompting \cite{zhou2023least}, further extend this principle by breaking down difficult tasks into a sequence of simpler sub-problems, thereby allowing LLMs to tackle tasks beyond their immediate capabilities.  

Beyond sequential prompting, researchers have proposed methods to improve robustness and reduce variance in LLM reasoning.  
Self-consistency \cite{wang2022self}, for example, samples multiple reasoning paths instead of relying on a single deterministic output, and then aggregates the results by majority voting or confidence scoring.  
This ensemble-like strategy mitigates errors caused by spurious reasoning chains and yields more reliable outputs across different problem domains.  
Other approaches such as instruction tuning \cite{ouyang2022training} \cite{wei2021finetuned} and reinforcement learning with human feedback (RLHF) \cite{lee2024instructpatentgpt} optimize prompts at the training stage, ensuring that models better follow instructions and align with user expectations at inference time.  

\subsection{External Information Integration.}
Another important direction emphasizes \textbf{external information integration}.  
Retrieval-augmented generation (RAG) \cite{wang2023retrieval} and related methods couple LLMs with information retrieval systems, allowing models to access knowledge bases or document collections during inference.  
By conditioning generation on retrieved passages, these methods enable LLMs to produce responses grounded in external evidence rather than relying only on parametric memory.  
This has proven especially valuable in dynamic domains such as open-domain question answering, where knowledge evolves rapidly and must be continuously updated.  

In addition to retrieval, hybrid systems combine statistical language models with symbolic or structured reasoning components.  
For example, neuro-symbolic approaches \cite{luo2025chatrule} augment LLMs with logic rules or knowledge graphs, allowing them to check consistency, enforce constraints, and perform structured reasoning tasks.  
Similarly, tool-augmented models \cite{schick2023toolformer} \cite{niketan2024integrating} train LLMs to call external APIs or computational tools when needed, enhancing their problem-solving flexibility.  
These approaches share the idea that LLMs alone are insufficient for robust reasoning; grounding them with external evidence or capabilities significantly improves factual accuracy, adaptability, and transparency.

\subsection{Difference-Based and Contrastive Reasoning.}
Recent works have also explored \textbf{difference-based or contrastive reasoning}, which highlights the importance of comparing inputs to detect changes, anomalies, or discriminative features.  
In representation learning, contrastive learning has become a dominant paradigm, where models are trained to bring similar examples closer and push dissimilar examples apart in the embedding space \cite{gao2021simcse}.  
This principle has been applied to improve sentence embeddings, multimodal alignment, and few-shot generalization, demonstrating the effectiveness of explicitly modeling differences.  
In the context of LLMs, contrastive prompting techniques have also been investigated, where the model is asked to compare multiple candidate outputs and justify the distinctions, thereby encouraging more careful and grounded reasoning.  

In vision–language reasoning, temporal and spatial differences have been studied for detecting events, changes, or anomalies in multimodal streams.  
For example, anomaly detection frameworks \cite{georgescu2021anomaly} utilize changes in visual features across time to flag abnormal patterns in surveillance or monitoring systems.  
Similarly, in video question answering and story understanding tasks, identifying what has changed between frames is essential for correct reasoning about events.  
While existing methods leverage differences for specific tasks like anomaly detection, they have not been used to enhance LLM reasoning. Our work addresses this gap by introducing a general framework that places differences at the core of the reasoning process, extending its applicability beyond visual domains.

\subsection{Overall Comparison.}
Taken together, existing research has advanced LLM reasoning through prompt design, retrieval augmentation, and difference-aware methods.  
Prompt engineering improves the ability of LLMs to follow reasoning steps, while external integration enhances their access to factual knowledge and computational tools.  
Contrastive approaches highlight the importance of recognizing differences, whether in text, vision, or multimodal contexts, as a means of improving robustness and grounding.  
However, these approaches remain somewhat fragmented, with each addressing a specific weakness of LLMs rather than providing a unified framework.  

Our work differs in that it systematically positions \textbf{differences across time and space as first-class signals} for reasoning.  
Rather than using differences as an auxiliary mechanism (e.g., for training embeddings or anomaly detection), we frame them as the starting point for reasoning itself, enabling LLMs to transition from passive answering to active question discovery.  
This distinction aligns our approach with human cognitive processes, where noticing change often triggers new inquiry.  
% By formalizing difference-guided reasoning and verifying it empirically, this paper contributes a novel direction that integrates insights from multiple lines of research into a coherent framework.  

\section{Difference-Guided Reasoning Model} \label{sec_model}

Large Language Models (LLMs) are effective at answering questions but often lack the ability to \textit{formulate new questions} on their own. Unlike human reasoning, which frequently begins by noticing a discrepancy or anomaly, LLMs tend to operate passively, generating answers only when explicitly prompted. To simulate more \textit{active thinking}, we propose a model that emphasizes identifying \textbf{differences}---in time or space---as the starting point for reasoning.  

The core idea is that \textbf{differences imply the need for new solutions}. By detecting differences, the system can both generate new questions and guide itself toward more relevant actions. This process enables LLMs to move closer to human-like reasoning, where recognizing ``what has changed'' is often the first step toward problem-solving.  

\subsection{Feature and Difference}
\textbf{Features} represent the defining characteristics of an object. Identifying these features is essential before deciding on actions, as it reduces interference from irrelevant factors. However, LLMs do not always prioritize feature recognition, which can lead to scattered or unfocused outputs. Detecting differences serves as an indirect but effective way of discovering features.  

Now we formally give the definition of difference. 
Let $x_i$ and $x_j$ denote two states (or representations) of the same object.  
We define the \textit{difference} between them as a mapping
\[
\Delta(x_i, x_j) = f(x_i) - f(x_j),
\]
where $f(\cdot)$ is a feature extractor that maps raw observations to a feature space.  

Key properties are:
\begin{itemize}
    \item $\Delta(x_i, x_j) = 0$ if the two states are indistinguishable in the chosen feature space.  
    \item $\|\Delta(x_i, x_j)\|$ quantifies the magnitude of change, which may be temporal (if $i,j$ represent different times) or spatial (if $i,j$ represent different sub-parts of the object).  
    \item The \textbf{impact} of a difference can be measured by a function $impact(\Delta(x_i, x_j)) \in \mathbb{R}_{\ge 0}$, which evaluates the significance of the detected change.  
\end{itemize}

% \paragraph{Example.}
Consider two images of the same road taken at different times. In $x_i$, the red car is $30\,\text{m}$ behind a leading vehicle, while in $x_j$ the distance has reduced to $15\,\text{m}$.  
If $f(\cdot)$ extracts the position of the red car relative to the vehicle ahead, then
\[
\Delta(x_i, x_j) = f(x_i) - f(x_j) = 30 - 15 = 15 \, .
\]
This positive difference indicates that the red car is getting closer to the vehicle in front. The magnitude $\|\Delta(x_i, x_j)\| = 15$ (meters) can then be evaluated by the impact function, which may suggest an appropriate action such as reducing speed to maintain safety.

\subsection{Types of Differences}

\subsubsection{Main $n$ Differences}
In practice, multiple differences may exist simultaneously across various features or components. Since not all differences are equally important, we introduce an \textit{impact function} that evaluates the significance of each detected difference:  

\begin{equation}
    I_i = impact(\Delta_i),
\end{equation}

where $\Delta_i$ denotes the $i$-th detected difference and $I_i \in \mathbb{R}_{\ge 0}$ is its associated impact score.  
To select the most influential changes, we choose the top-$n$ differences according to their impact:  

\begin{equation} \label{eq_n_max}
    \mathcal{D}_{main} = \arg\max_{\,\mathcal{S} \subseteq \{\Delta_i\},\,|\mathcal{S}|=n} \sum_{\Delta_i \in \mathcal{S}} I_i .
\end{equation}

Here, $\mathcal{D}_{main}$ represents the set of \textit{main differences}. The choice of $n$ depends on the scenario and can be fixed by design or adaptively selected by the LLM, ensuring focus on the most critical changes without overwhelming the reasoning process.  

\subsubsection{Latest Difference}
While main differences capture broader patterns, the \textbf{latest difference}---the most recent change observed---is often decisive for immediate actions.  
Formally, let $t$ denote the current time and $\Delta_t$ the difference between states $(x_{t-1}, x_t)$. We define the latest difference as  

\begin{equation}
    \Delta_{latest} = \Delta(x_{t-1}, x_t).
\end{equation}

This value highlights the most recent change and is particularly useful for short-term decision-making, such as triggering a control adjustment in response to sudden environmental variation.  

The \textit{main differences} guide long-term or strategic reasoning by identifying the most influential changes, whereas the \textit{latest difference} directs immediate responses to the most recent event. Together, they provide a balanced mechanism for prioritizing actions.  

\subsection{Temporal Comparison}
\textbf{Temporal comparison} involves analyzing an object’s state across two or more time points in order to capture its evolution.  
Let the sequence of times be $t_0, t_1, \dots, t_k$. If the object is in state $s_{t_0}$ at time $t_0$ and $s_{t_1}$ at time $t_1$, then the difference between these two states is defined as  

\begin{equation}
    \delta_{t_0 \rightarrow t_1} = s_{t_1} - s_{t_0}.
\end{equation}

\begin{figure}[h]
    \centering
    \includegraphics[width = 3.5in]{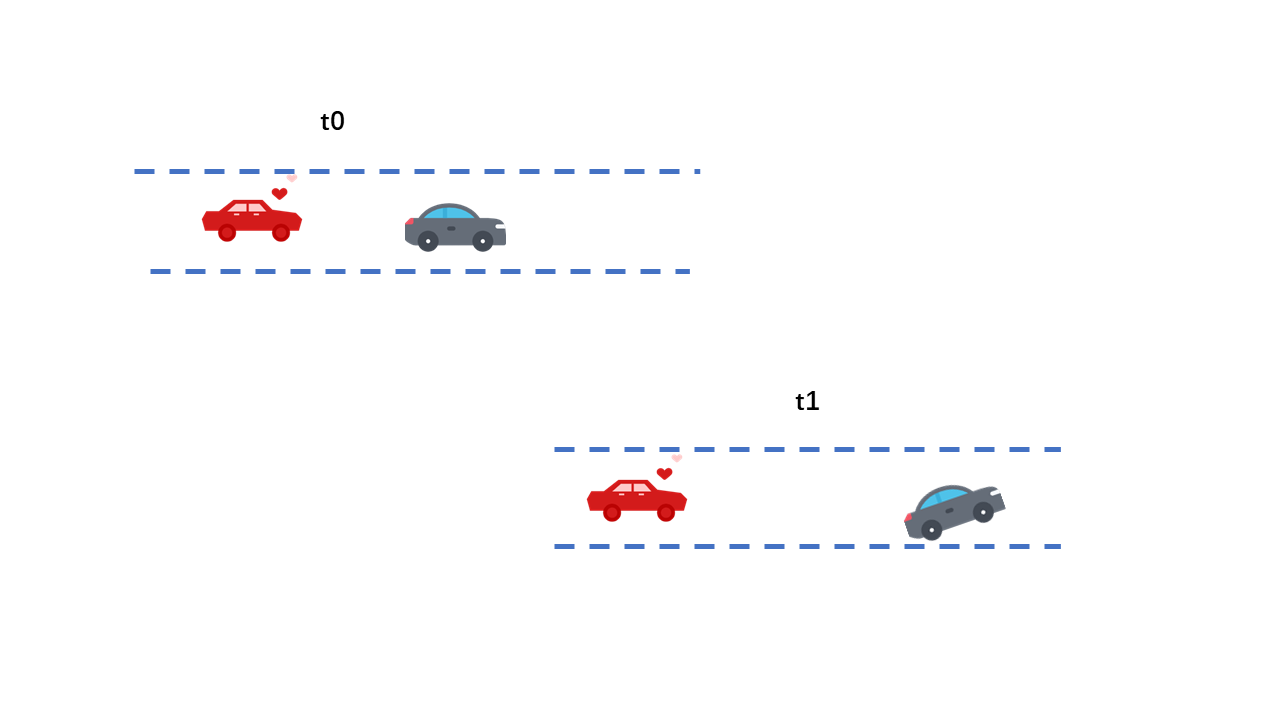}
    \caption{Illustration of temporal-based comparison.}
    \label{fig_time}
\end{figure}

More generally, a temporal difference operator can be written as  
\begin{equation}
    \delta_{t_i \rightarrow t_j} = f(s_{t_j}) - f(s_{t_i}), \qquad t_j > t_i,
\end{equation}
where $f(\cdot)$ extracts relevant features from the state representation.  
The resulting $\delta_{t_i \rightarrow t_j}$ highlights how the object changes over time and can be used to guide reasoning or actions (Figure~\ref{fig_time}).

\subsubsection{Compared with History}
When comparisons involve past data, we refer to this as a \textbf{comparison with history}.  
Formally, let $\mathcal{H} = \{s_{t_0}, s_{t_1}, \dots, s_{t_{k-1}}\}$ denote the set of stored historical states.  
A comparison at time $t_k$ can be defined as  
\begin{equation}
    \delta_{t_k} = s_{t_k} - \hat{s}, \qquad \hat{s} \in \mathcal{H},
\end{equation}
where $\hat{s}$ is a selected reference state (e.g., the most recent past, the average of prior states, or a prototype).  

A key design decision is how to \textbf{store historical states}. Instead of saving only simplified summaries, we advocate storing \textbf{full information} (raw data) and linking it to derived states. This ensures that when an action derived from $\delta_{t_k}$ proves inadequate, the system can revisit complete historical data for further analysis. Although this strategy requires more resources, it significantly improves traceability and robustness in decision-making.  

\subsection{Spatial Comparison}
\textbf{Spatial comparison} analyzes whether subcomponents of a larger object exhibit differences in their distribution or structure.  
Intuitively, a complex object (e.g., an image, a map, or a system) can be decomposed into smaller parts, which are then compared.  

Formally, let an object $O$ be partitioned into $m$ sub-objects:
\[
O \;\rightarrow\; \{ s_1, s_2, \dots, s_m \}.
\]
For each sub-object $s_i$, a feature representation is extracted:
\[
z_i = f(s_i),
\]
where $f(\cdot)$ denotes a feature extractor (e.g., measuring size, density, or position).  
Spatial differences can then be defined pairwise as
\[
\Delta_{ij} = z_i - z_j, \qquad i \neq j,
\]
and the overall spatial variability of $O$ may be summarized as
\[
V(O) = \frac{1}{m(m-1)} \sum_{i \neq j} \| \Delta_{ij} \|.
\]

\paragraph{Practical considerations.}
In practice, directly partitioning complex objects is often difficult, especially when the correct partitioning depends on context.  
A pragmatic approach is to prompt the LLM to \textbf{summarize distributions} across different regions, thereby capturing relevant differences without requiring strict predefined segmentation.  

\begin{figure}[h]
    \centering
    \includegraphics[width = 3.5in]{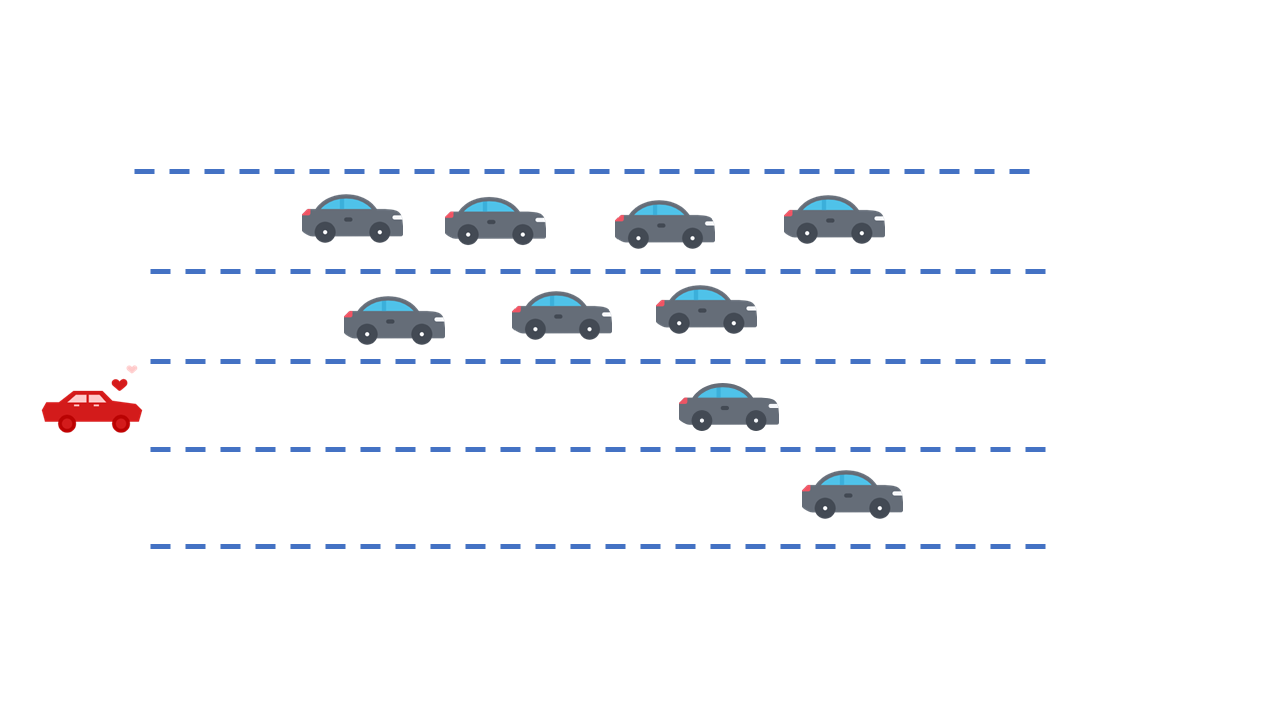}
    \caption{Illustration of spatial comparison in a road scene.}
    \label{fig_spatial}
\end{figure}

\paragraph{Example.}
Consider a driving scene on a road. Some sections of the road are wider than others, leading to variations in lane width and vehicle density (Figure~\ref{fig_spatial}).  
If we let $s_1, s_2, s_3$ denote three different road segments, then $z_i = f(s_i)$ could represent the average lane width or traffic density in segment $s_i$.  
Spatial comparison highlights the differences $\Delta_{ij}$ between these segments, enabling the system to recognize that certain parts of the road may require different driving strategies, such as reducing speed in narrower segments or preparing for congestion in denser ones.  

\subsection{Abnormal Behavior}
Differences between states can be either small (normal variations) or large (potential anomalies).  
A state is regarded as \textbf{abnormal} when the magnitude of its change exceeds a predefined threshold, as expressed in \eqref{eq_abnormal}:

\begin{equation} \label{eq_abnormal}
    \| \Delta \| > \theta ,
\end{equation}
where $\Delta$ denotes the observed difference between states and $\theta$ is the threshold.  

\paragraph{Threshold-based detection.}
The threshold $\theta$ is typically estimated from historical data, for example by taking statistical bounds (e.g., mean $\pm$ standard deviation) of normal variations.  

\paragraph{History-based detection.}
In many cases, however, no clear threshold is available. Instead, detection relies on comparing the current state $s_t$ with a set of previously observed (historical) states:
\[
s_t \;\;\text{is abnormal if}\;\; d(s_t, \mathcal{H}) > \eta,
\]
where $\mathcal{H}$ denotes the historical memory of normal states, $d(\cdot,\cdot)$ is a distance or dissimilarity measure, and $\eta$ is a tolerance level.  
If $s_t$ is significantly different from all historical states, it is flagged as abnormal.

\paragraph{Representation of normality.}
Normal and abnormal behaviors can be recorded as pairs $(s, a)$, where $s$ denotes the state and $a \in \{\text{normal}, \text{abnormal}\}$ indicates its status.  
The historical memory $\mathcal{H}$ thus serves as a repository of common patterns, against which new states are evaluated.  
When a new difference $\Delta$ exceeds the variability represented in $\mathcal{H}$, the system concludes that an abnormal behavior has occurred and selects an appropriate response.  

\subsection{External Information for Detecting Differences}
Detecting differences does not rely solely on the pre-trained knowledge of an LLM.  
In many situations, \textbf{external information} from users, databases, or sensors must be incorporated to expose changes that are not evident from the model’s internal knowledge alone.  

Let $s_t$ denote the current state of an object at time $t$.  
If only the LLM’s internal knowledge is used, differences can be computed as
\[
\Delta_t^{int} = f(s_t) - f(s_{t-1}),
\]
where $f(\cdot)$ extracts relevant features.  

However, richer detection is possible when external information $E_t$ is available (e.g., logs, sensor readings, or user-provided context).  
In this case, the enhanced difference is
\[
\Delta_t^{ext} = g(s_t, E_t) - g(s_{t-1}, E_{t-1}),
\]
where $g(\cdot)$ fuses the observed state with external evidence.  

\paragraph{Examples.}
\begin{itemize}
    \item Providing historical snapshots of an object (e.g., archived sensor data) enables detection of long-term drifts.  
    \item Adding user-provided metadata (e.g., environmental conditions) helps explain why certain differences occur.  
    \item Combining multiple sensor modalities (e.g., camera + LiDAR) produces a more reliable estimate of change than either source alone.  
\end{itemize}

\paragraph{Implication.}
By integrating external information, the system can (i) detect subtle differences that internal knowledge might miss, and (ii) ground its reasoning in concrete, real-world evidence.  
This makes the process of difference detection more robust and reduces the risk of overlooking meaningful changes.  

\section{Verification} \label{sec_verification}
The effectiveness of the proposed difference-guided reasoning framework is evaluated through experiments in two scenarios: \textbf{temporal difference} and \textbf{spatial difference}.  
The objectives of verification are twofold:  
\begin{enumerate}
    \item to demonstrate that explicitly prompting differences enables the LLM to focus on critical changes and produce more accurate reasoning, and  
    \item to show that emphasizing differences reduces redundant information by concentrating only on the changed states.  
\end{enumerate}

For all experiments, \textbf{ChatGPT-5} is used as the LLM platform.  

\subsection{temporal difference}
This experiment considers two sequential images of a red car driving on a road.  
The two pictures are largely similar, but the second image shows the red car closer to the vehicle ahead, with more vehicles visible in the scene.  
The objective is to evaluate whether explicitly prompting the LLM with differences helps it focus on the key issue: the need to reduce speed to maintain a safe distance.  

\begin{figure}[h]
    \centering
    \includegraphics[width = 3.5in]{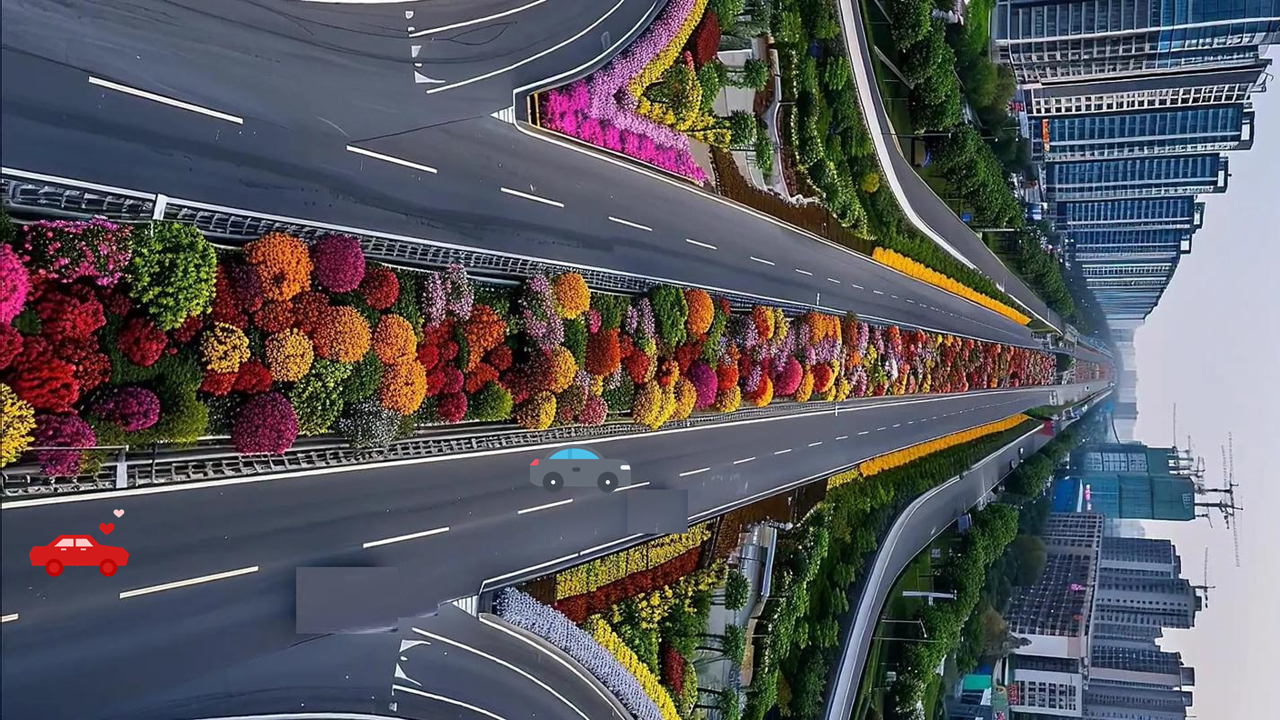}
    \caption{The first picture in the time sequence.}
    \label{fig_t0}
\end{figure}

\begin{figure}[h]
    \centering
    \includegraphics[width = 3.5in]{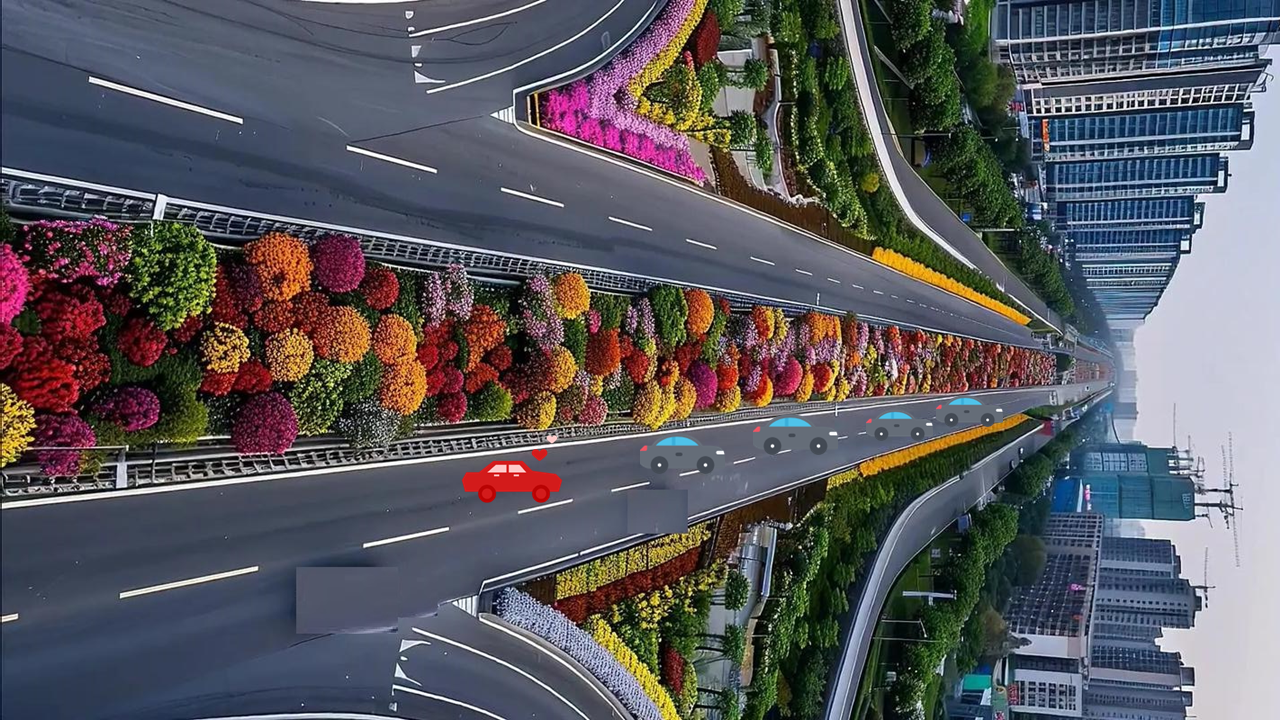}
    \caption{The second picture in the time sequence.}
    \label{fig_t1}
\end{figure}

\paragraph{Procedure.}
The verification process consists of two steps:  
\begin{enumerate}
    \item The two images are uploaded and introduced with the neutral prompt:  
    \textit{``The above are two pictures. Please do not reply now; reply when required later. Is this OK?''}  
    The images are shown in Figures~\ref{fig_t0} and \ref{fig_t1}.  
    \item Two prompting strategies are compared:  
    \begin{itemize}
        \item \textbf{Direct Method}: The LLM is asked \textit{``How do you think of this picture?''} without being explicitly directed to identify differences.  
        \item \textbf{Difference Method}: The LLM is first prompted with \textit{``These two images depict events that occurred sequentially. What is the difference between the two pictures?''} before reasoning about the scene.  
    \end{itemize}
\end{enumerate}

\paragraph{Evaluation.}
Each method is repeated 20 times.  
Generated responses are compared with a reference statement:  
\begin{quote}
``The distance between the red car and the vehicle ahead is decreasing, and there are more vehicles in front.  
It is recommended to slow down immediately to maintain a safe following distance.''
\end{quote}
We compute the \textbf{cosine similarity} between response embeddings and the reference embedding using the \texttt{sentence\_transformers} library:  
\[
\text{sim}(y, y^*) = \frac{\langle e(y), e(y^*) \rangle}{\| e(y) \| \, \| e(y^*) \|}.
\]

\begin{figure*}[t]
    \centering
    \includegraphics[width = 5.5in]{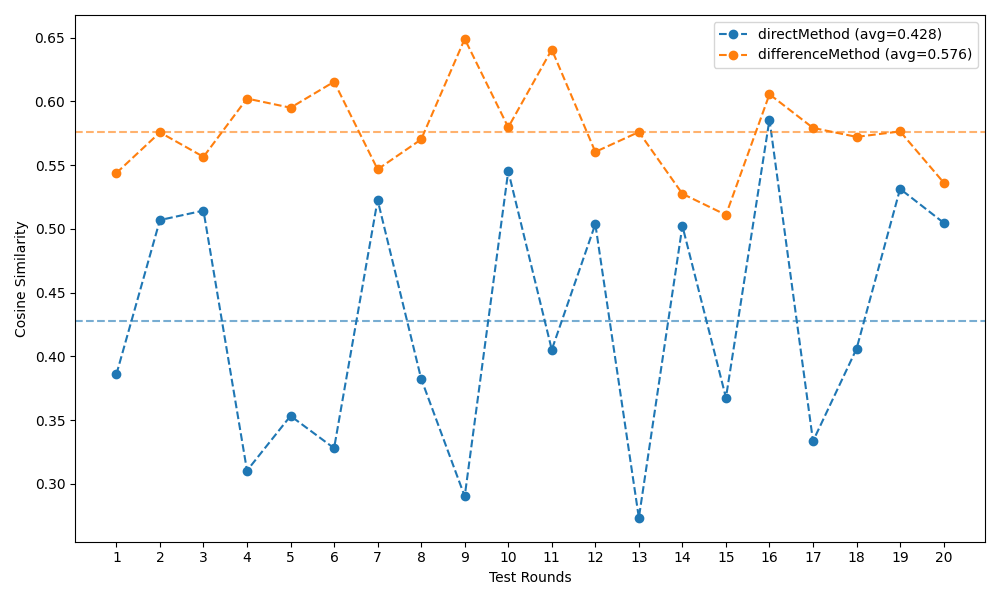}
    \caption{Comparison of Direct Method and Difference Method in temporal-difference reasoning.}
    \label{fig_timeDiffernenceResults}
\end{figure*}

\paragraph{Results.}
The results are shown in Figure~\ref{fig_timeDiffernenceResults}.  
The Difference Method achieves a higher average similarity ($\mu_\Delta = 0.5760$) than the Direct Method ($\mu_d = 0.4276$).  
The maximum similarity for the Difference Method is $0.6491$ (9th trial) and the minimum is $0.5110$ (15th trial).  
In contrast, the Direct Method reaches a maximum of $0.5858$ (16th trial) but drops as low as $0.2733$ (13th trial).  
Although the Direct Method occasionally aligns with the key issue, its focus is inconsistent and appears random.  
By contrast, the Difference Method reliably emphasizes the relevant factor—maintaining safe distance—over 20 trials.  

\paragraph{Analysis.}
Qualitative inspection confirms that responses from the Difference Method often recommend slowing down or maintaining distance, while Direct Method outputs frequently describe irrelevant aspects such as background scenery.  
This suggests that prompting with differences suppresses distractions and helps the LLM focus on the safety-critical issue.  

A two-sample $t$-test further validates the findings.  
The computed $t$-value exceeds the critical threshold at the $0.05$ significance level, with $p < 0.01$.  
Thus, we reject the null hypothesis $H_0 : \mu_d = \mu_\Delta$ and conclude that the Difference Method provides a statistically significant improvement over the Direct Method.  

\begin{figure}[h]
    \centering
    \includegraphics[width=4in]{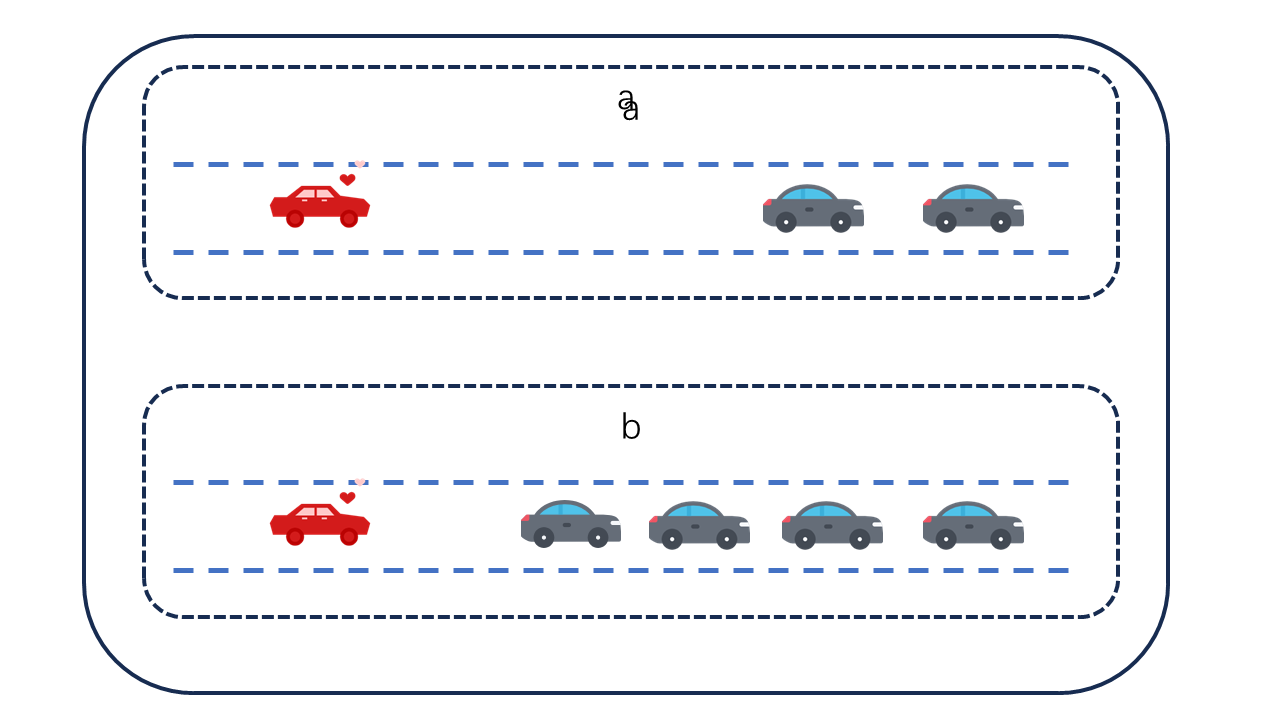}
    \caption{Simplified traffic scene with only lane borders and cars, used to test LLM focus without background distraction.}
    \label{fig_time_verification_simple}
\end{figure}

In fact, when we remove the background from the original images (Figure~\ref{fig_t0} and Figure~\ref{fig_t1}) and retain only the essential traffic elements---cars and simple lane borders, as shown in parts (a) and (b) of Figure~\ref{fig_time_verification_simple}---the LLM is able to directly compare and analyze the traffic situation.  
This behavior is similar to human reasoning: when the environment is simple, attention naturally focuses on the core issue.  
However, in complex scenarios with many distracting details, explicitly highlighting differences becomes essential for guiding the LLM to concentrate on the most relevant aspects.  

\subsection{Spatial Difference}
This experiment considers a hand-drawn schematic of a metro train, commonly used for airport transportation.  
Unlike conventional metro trains, where all carriages are of equal length, this train has carriages of varying sizes.  
The variation arises from different seating layouts: some cars have forward-facing rows, while others have face-to-face seating, resulting in unequal interior space.  
The objective is to verify whether the LLM can identify this irregularity and explain the underlying reason.  

\begin{figure}[h]
    \centering
    \includegraphics[width = 3.5in]{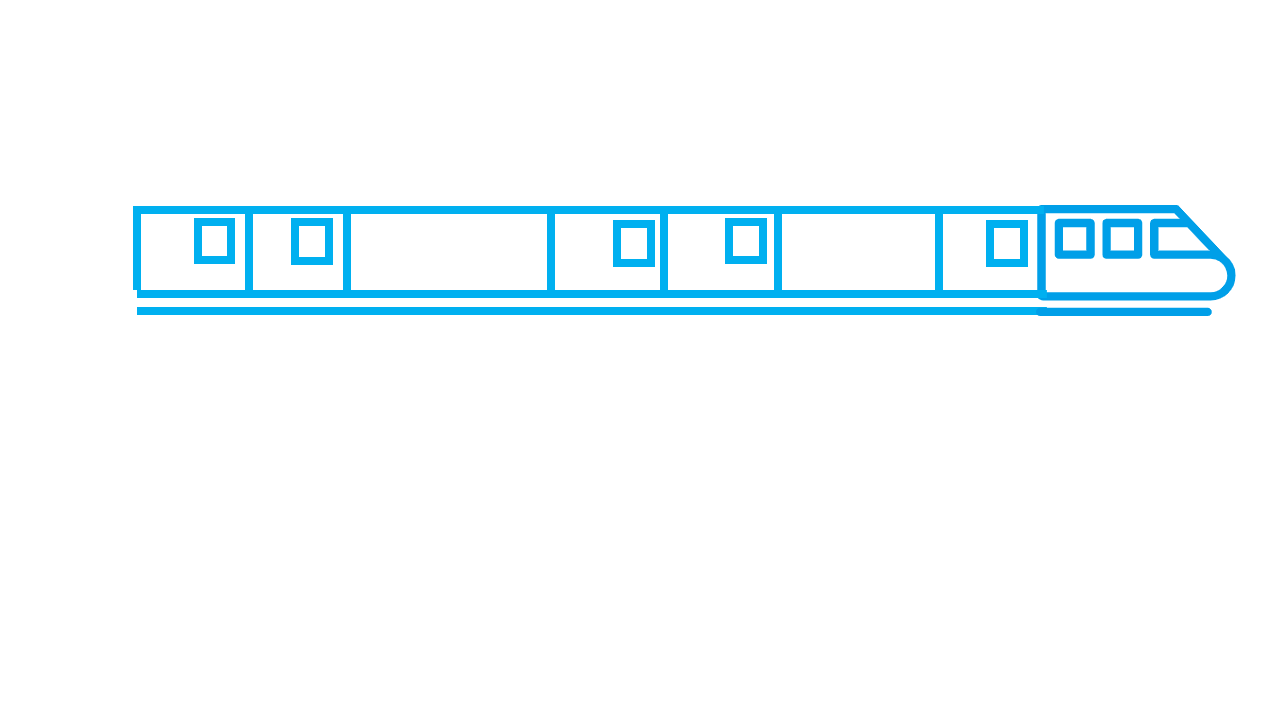}
    \caption{Hand-drawn schematic of a metro train with unequal carriage lengths.}
    \label{fig_position_differnence_v1}
\end{figure}

\paragraph{Procedure.}
The verification process is conducted in two steps:  
\begin{enumerate}
    \item The train image is uploaded with the neutral prompt:  
    \textit{``Please do not reply now and reply only when required later. Is this OK?''}  
    The schematic is shown in Figure~\ref{fig_position_differnence_v1}.  
    \item Two prompting strategies are compared:  
    \begin{itemize}
        \item \textbf{Direct Method}: The LLM is asked \textit{``How do you think of this picture?''}, without directing attention to specific differences.  
        \item \textbf{Difference Method}: The LLM is explicitly asked \textit{``Is there some difference in the carriages? If yes, why is there a difference?''}.  
    \end{itemize}
\end{enumerate}

\paragraph{Evaluation.}
Each method is executed 20 times.  
Responses are compared with a reference description:  
\begin{quote}
``The carriages of the train vary in size, primarily due to the non-uniform seating layout across different cars.  
Some carriages are arranged with forward-facing rows, while others feature face-to-face seating, resulting in differences in interior space.  
This type of train is commonly used in airport metro lines.''
\end{quote}

Cosine similarity between generated responses and the reference is computed using the \texttt{sentence\_transformers} library.  

\begin{figure*}[t]
    \centering
    \includegraphics[width = 5.5in]{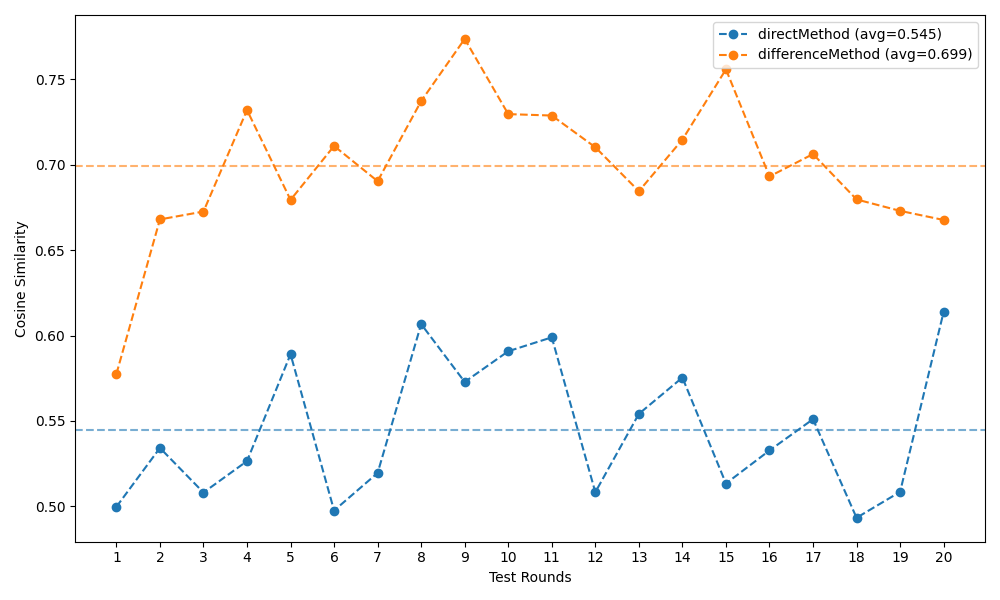}
    \caption{Comparison of Direct Method and Difference Method in spatial-difference reasoning.}
    \label{fig_positionDiffernenceResults}
\end{figure*}

\paragraph{Results.}
The results, shown in Figure~\ref{fig_positionDiffernenceResults}, indicate that the Difference Method consistently outperforms the Direct Method.  
The average similarity of the Difference Method is $\mu_\Delta = 0.6992$, compared to $\mu_d = 0.5448$ for the Direct Method.  
The Difference Method reaches a maximum similarity of $0.7735$ (9th trial) and a minimum of $0.5775$ (1st trial).  
Except for the first run, all results from the Difference Method are higher than those of the Direct Method.  

Qualitative analysis shows that Direct Method outputs often fail to note the carriage differences, focusing instead on irrelevant details.  
By contrast, Difference Method outputs consistently describe the unequal carriage lengths and attribute them to differences in seating design.  
This suggests that explicitly prompting differences encourages the LLM to engage in a more \textbf{active reasoning process}, identifying novel and meaningful observations rather than passively describing the input.

A two-sample $t$-test further confirms the improvement.  
With $\mu_\Delta = 0.6992$ and $\mu_d = 0.5448$, the $t$-test result shows a statistically significant difference ($p < 0.01$).  
Thus, we reject the null hypothesis and conclude that the Difference Method yields more accurate and consistent reasoning in spatial comparison tasks.

\section{Conclusion} \label{sec_conclusion}
In this paper, we proposed a \textbf{difference-guided reasoning framework} to enhance the active reasoning capability of Large Language Models (LLMs).  
The framework formalizes differences across time and space, prioritizes the most impactful and latest changes, and links them to corresponding actions.  
We further extended the model with abnormal behavior detection and the integration of external information, allowing the system to recognize subtle variations and respond more effectively.  
Verification experiments on temporal-difference and spatial-difference tasks demonstrated that explicitly prompting for differences improves both focus and reasoning accuracy.  
Statistical analysis confirmed that the Difference Method consistently outperforms the Direct Method, highlighting the value of difference-guided reasoning.  

Future work will explore several directions.  
First, we aim to generalize the framework across more complex multimodal tasks, such as combining video, sensor, and textual data.  
Second, adaptive methods for automatically selecting the most relevant differences and dynamically adjusting thresholds for abnormal behavior detection will be investigated.  
Finally, we plan to extend difference-guided reasoning to collaborative multi-agent settings, where multiple LLMs or LLM–sensor systems work together to discover, share, and act upon differences in real time.

\ifCLASSOPTIONcaptionsoff
  \newpage
\fi

\bibliographystyle{IEEEtran}
\bibliography{ref}


\begin{IEEEbiography}{Hong Su}
  received the MS and PhD degrees, in 2006 and 2022, respectively, from Sichuan University, Chengdu, China. He is currently a researcher of Chengdu University of Information Technology Chengdu, China. His research interests include blockchain, cross-chain and smart contract.
\end{IEEEbiography}


%\vfill


\end{document}